\documentclass{article}
\usepackage{graphicx} 
\usepackage{color}
\usepackage{tikz-cd}
\usepackage{url}
\usepackage{soul}
\usepackage[nottoc]{tocbibind}
\usepackage{amsmath}
\usepackage{mathtools}
\usepackage{amssymb}
\usepackage[svgnames]{xcolor}
\usepackage{braket}
\usepackage[utf8]{inputenc}
\usepackage{tikz}
\usetikzlibrary{backgrounds,calc,positioning}
\graphicspath{ {./images/} }
\usepackage{natbib}
\usepackage{authblk}
\usepackage{tcolorbox}

\title{Origin of the Covariant Wigner Operator\\ as a Quantum Amplitude in QCD}

\author{Chueng-Ryong Ji$^\dagger$ and Daniel W. Piasecki\\
Department of Physics, Box 8202,\\ North Carolina State University, Raleigh, NC 27695, USA\\ }
\date{January 2026}
\begin{document}

\maketitle

\section*{Abstract}

The Wigner function plays a central role in QCD as a phase space object encoding correlations among quarks, antiquarks, and gluons, yet its interpretation remains subtle due to its quasiprobabilistic nature and possible negativity. Recent work based on the Koopman–von Neumann–Sudarshan (KvNS) Hilbert space formulation of classical mechanics suggests the Wigner function arises as a \textit{quantum probability amplitude} projected onto classical phase space, rather than a quasiprobability density \citep{wignerphasespace, wave_operator}. In the classical limit, this amplitude reduces to the classical Koopman wavefunction. In this work, we extend this perspective to relativistic QCD by constructing a Koopman description of the quark Wigner operator. We show that the Wigner operator is naturally isomorphic to a phase space spinor, providing a unified framework in which both classical and quantum dynamics are expressed. Within this formulation, the Wigner function retains its interpretation as an amplitude even in the relativistic regime. This viewpoint clarifies the origin of negativity and other nonclassical features, and provides a more transparent foundation for parton distribution functions in QCD. Remarkably, the relativistic Koopman framework reproduces the classical limit of QCD.\newline\newline\newline
$\dagger$ Corresponding Author\\

\section{Introduction}

While the Wigner operator is widely employed in QCD to probe correlations among quarks, antiquarks, and gluons, its dependence on both position and momentum variables appears puzzling in light of the uncertainty principle, which forbids their simultaneous specification in the Hilbert space formulation of quantum mechanics. Moreover, the probabilistic interpretation of the Wigner function is limited as it can be negative, although the integration of either position or momentum variables leads to the positive-definiteness of the Wigner function. To clarify these features, we turn to a Hilbert space formulation of \textit{classical} dynamics, corresponding to the formal limit $\hbar \to 0$, and motivate this perspective through the structure of QCD correlation functions.

Koopman-von Neumann-Sudarshan (KvNS) mechanics is a branch of classical statistical mechanics that has recently experienced a revival of interest since its inception by \citet{koopman}, \citet{vN}, and \citet{Sudarshan}. It is a fully \textit{classical} theory completely based in the \textit{Dirac-von Neumann axioms} and a complex Hilbert phase space with a classical wavefunction, Born Rule, operators, and commutator structure \citep{kvnwaves,MauroPhDThesis,ODM,relODM,McCauletal,PiaseckiThesis}. It has led to numerous avenues of investigation, including quantum-classical system coupling \citep{QM_CM_hybrids_2,Koopmanwavefunctions, QM_CM_hybrids_1}, irreversibility of dynamics and origin of stochasticity from determinism \citep{McCauletal}, quantum friction \citep{qfriction1,qfriction2a,qfriction2b,qfriction3,qfriction4}, a new unified representation of classical-quantum dynamics \citep{wave_operator}, speed limit for quantum dynamics \citep{qspeedlimit}, novel generalizations of measurement theory \citep{Morgan}, advancement of efficient numerical methods \citep{algoefficiency1,OpenDirac,algoefficiency2,algoefficiency3}, and inspired novel treatments of classical electromagnetism \citep{CE16, DvN_EM}. Perhaps most importantly, the KvNS approach has led to the paradigm of Operational Dynamic Modeling (ODM), which melds together quantum algebras with classical algebras in a universal framework for theoretical investigation \citep{ODM,wignerphasespace, relODM}. 

A principle discovery of the ODM approach is a new interpretation of the famous Wigner quasiprobability distribution,
\begin{equation}\label{eq: wigner_function}
    W(x,p) = \frac{1}{2\pi} \int d\lambda_p ~\rho\Big(x - \frac{\hbar}{2}\lambda_p, x + \frac{\hbar}{2}\lambda_p\Big)e^{ip\lambda_p},
\end{equation}
as a quantum probability amplitude projected into a point of classical phase space \citep{ODM, wignerphasespace, OpenDirac,relODM}. In other words, in the limit of $\hbar \rightarrow 0$, the Wigner function becomes a KvNS classical wavefunction amplitude \citep{ODM,wignerphasespace}, which does not need to remain positive definite, unlike a true probability distribution.
The Wigner function, of course, has a central place in the heart of quantum theory \citep{wigner_general}. In QCD, it is also important for understanding the properties of quarks and gluons \citep{Quark_Distribution_9,quark_dist_book,Quark_Distribution_11,Wigner_Operator_extra_reference,GPDs, Quark_Distributions_2,Quark_Distributions_3, Quark_Distributions_1, Quark_Distributions_4,quark_wigner_2017,gluon_wigner_1,quark_wigner_2018,wigner_phd_thesis,Quark_Distribution_5,Quark_Distribution_7,Quark_Distribution_6}.
Here, we discuss an extension of ODM arguments to derive the Wigner Operator as the root of Parton Distribution Functions (PDFs), Generalized Parton Distributions (GPDs), and Transverse Momentum Dependent Parton Distributions (TMDs) for quarks and gluons. This gives us a fundamentally clearer picture to the origin of the PDFs, GPDs, and TMDs, instead of postulating their mathematical root, the Wigner function, \textit{a priori}. 

\subsection{Non-Relativistic ODM Theory of the Wigner Distribution: Schr{\"o}dinger field}\label{nonrelativistic_section}

ODM is a universal theoretical framework for creating dynamical models from collected observations and data \citep{ODM,relODM}. Important ingredients for ODM include  the definitions
of the observables’ average, and the algebra of the
observables. We additionally utilize Ehrenfest-like theorems that describe the time evolution of the average observables for reconstructing the dynamical model. 

For deriving the Wigner distribution for the non-relativistic Schr{\"o}dinger field, we will begin by utilizing ODM with a unified quantum-classical basis in the Dirac-von Neumann axioms. Classical physics may have an underpinning in the Dirac-von Neumann axioms commonly attributed to quantum theory through the complex Hilbert Space approach of \citet{koopman}, \citet{vN}, and \citet{Sudarshan}. 
The Dirac-von Neumann axioms are briefly summarized \citep{dirac,von_N_textbook}:

\begin{enumerate}
    \item The 
    wavefunction ket $\ket{\psi}$ (a vector in complex Hilbert Space) describes the state of the system
    \item For observable $\zeta$, there is an associated Hermitian operator $\hat{\zeta}$ which obeys the eigenvalue problem $\hat{\zeta}\ket{\zeta} = \zeta\ket{\zeta}$, where $\zeta$ is the value seen by measurement. Two common observables are position $\hat{x}$ and momentum $\hat{p}$, which obey (in one dimension),

    \begin{equation}
        \hat{x}\ket{x}=x\ket{x},
    \end{equation}
    \begin{equation}
        \hat{p}\ket{p}=p\ket{p}.
    \end{equation}
    \item Born Rule: The probability density for making any measurement for observable $\zeta$ is given by 
    \begin{equation}
        \rho_\zeta = \braket{\psi|\zeta}\braket{\zeta|\psi}.
    \end{equation}
    Upon measurement, the state of the system collapses from $\ket{\psi}$ to $\ket{\zeta}$.
    
    \item The state space of a composite system is the tensor product of the subsystem’s state spaces, $\mathcal{H} = \mathcal{H}_1 \otimes \mathcal{H}_2 \otimes ...$

    \item Unitary time evolution: \begin{equation}\label{eq: stones_theorem}
    \ket{\psi(t)}=\hat{U}(t-t')\ket{\psi(t')}
\end{equation}
where
\begin{equation}\label{eq:propogator}
    \hat{U}=\exp(-i\hat{\Omega}(t-t')).
\end{equation}
%
%
The operator $\hat{\Omega}$ is Hermitian and here the infinitesimal generator of time evolution of the wavefunction.
\end{enumerate}
These are a universal set of axioms describing \textit{both classical and quantum systems} \citep{kvnwaves,MauroPhDThesis,ODM,relODM,McCauletal,PiaseckiThesis}.

The way we distinguish between classical and quantum systems is through their respective algebras.
For classical mechanics the position and momentum will commute,

\begin{equation}\label{eq: classical_comm_1a}
    [\hat{x},\hat{p}]=0,
\end{equation}
corresponding to the $\hbar \rightarrow 0$ limit.
Additionally, for classical mechanics we postulate new classical operators $\hat{\lambda}_x$ and $\hat{\lambda}_p$ which obey the commutation rules
\begin{equation}\label{eq: koopman_algebra}
    [\hat{x},\hat{\lambda}_x]=[\hat{p},\hat{\lambda}_p] = i.
\end{equation}
These new operators are termed the ``hidden dynamical variables" of classical mechanics \citep{Sudarshan}. These auxiliary operators can be viewed as a class of Lagrange multipliers enforcing classical dynamics \citep{McCauletal}. These will be shown to be directly related to the already familiar Bopp operators \citep{bopp}. 

Using the Dirac-von Neumann axioms underpinning the KvNS theory and unitary time evolution (Stone's theorem) along with the above classical algebra, one can recover the Liouville equation and the Poisson bracket. Starting from the Ehrenfest theorems,
\begin{equation}\label{eq: ehrenfest_thm_1}
    \frac{d}{dt}\bra{\psi(t)}\hat{x}\ket{\psi(t)}=\bra{\psi(t)}\frac{\hat{p}}{m}\ket{\psi(t)},
\end{equation}
\begin{equation}\label{eq: ehrenfest_thm_2}
    \frac{d}{dt}\bra{\psi(t)}\hat{p}\ket{\psi(t)}=-\bra{\psi(t)}U'(\hat{x})\ket{\psi(t)},
\end{equation}
we differentiate and utilize Stone's theorem (eq. \ref{eq: stones_theorem}) to get

\begin{equation}
    \bra{\psi(t)}\hat{x}\hat{\Omega}-\hat{\Omega}\hat{x}\ket{\psi(t)}  = \bra{\psi(t)}\frac{i\hat{p}}{m}\ket{\psi(t)},
\end{equation}
\begin{equation}
\bra{\psi(t)}\hat{p}\hat{\Omega}-\hat{\Omega}\hat{p}\ket{\psi(t)}  = \bra{\psi(t)}-iU'(\hat{x})\ket{\psi(t)}.
\end{equation}
Dropping the wavefunction bra and kets from each side, the commutation relationships that result can be used with the classical algebra to figure out the form of the classical generator of motion. Here, the importance of the auxiliary, new classical operators comes into play. If we only use commuting $\hat{x}$ and $\hat{p}$ operators for the classical algebra, we run into a contradiction:

\begin{equation}\label{eq: contra_1}
    [\hat{x},\hat{\Omega}(\hat{x},\hat{p})] = \frac{i\hat{p}}{m} ~~~~\Rightarrow ~~~~ p = 0,
\end{equation}
\begin{equation}\label{eq: contra_2}
    [\hat{p},\hat{\Omega}(\hat{x},\hat{p})] = -iU'(\hat{x}) ~~~~\Rightarrow ~~~~ U'(x) = 0.
\end{equation}
This clearly excludes many physical situations of interest. With the previous Koopman classical algebra (eq. \ref{eq: koopman_algebra}), we circumvent this issue:
\begin{equation}
    [\hat{x},\hat{\Omega}(\hat{x},\hat{p}, \hat{\lambda}_x,\hat{\lambda}_p)] = \frac{i\hat{p}}{m} ~~~~\Rightarrow ~~~~ [\hat{x},\hat{\lambda}_x]\frac{\partial \hat{\Omega}}{\partial \lambda_x}=\frac{i\hat{p}}{m}  ~~~~\Rightarrow ~~~~ \hat{\Omega} = \frac{\hat{p}\hat{\lambda}_x}{m} + C,
\end{equation}
\begin{equation}
    [\hat{p},\hat{\Omega}(\hat{x},\hat{p}, \hat{\lambda}_x,\hat{\lambda}_p))] = -iU'(\hat{x}) ~~~~\Rightarrow ~~~~ [\hat{p},\hat{\lambda}_p]\frac{\partial \hat{\Omega}}{\partial \lambda_p} = -iU'(\hat{x}) ~~~~\Rightarrow ~~~~ \hat{\Omega} = -U'(\hat{x})\hat{\lambda}_p+ C.
\end{equation}
This is a derivation of the Koopman operator, which we summarize as (relabeling as $\hat{K}$):

\begin{equation}
    \hat{K} = \frac{\hat{p}\hat{\lambda}_x}{m}  -U'(\hat{x})\hat{\lambda}_p+ C(\hat{x},\hat{p}).
\end{equation}
To derive the Liouville equation, we start with Stone's theorem again, sandwiching from the left with a bra of classical (commuting) position and momentum:

\begin{equation}
    \ket{\dot{\psi}} = -i\hat{K}\ket{\psi},
\end{equation}
\begin{equation}\label{eq: int_step}
    \frac{\partial}{\partial t} \braket{x,p|\psi(t)} = -i\bra{x,p}\hat{K}\ket{\psi(t)} = \Big(\frac{p}{m}\frac{\partial}{\partial x}-\frac{\partial U}{\partial x}\frac{\partial }{\partial p}\Big) \braket{x,p|\psi(t)}.
\end{equation}
Using the Born rule (axiom 3) for classical wavefunctions, 

\begin{equation}
    \rho = \braket{\psi(t)|x,p}\braket{x,p|\psi(t)},
\end{equation}
we can rewrite eq. \ref{eq: int_step} as the famous Liouville equation:

\begin{equation}
     \frac{\partial\rho}{\partial t}   = \Big(\frac{p}{m}\frac{\partial}{\partial x}-\frac{\partial U}{\partial x}\frac{\partial }{\partial p}\Big) \rho = \{\rho,H\},
\end{equation}
where $H$ here is the classical Hamiltonian. An equivalent way to derive the Poisson bracket is through the Heisenberg equation of motion with the classical Koopman generator:

\begin{equation}
    \frac{d \hat{O}}{dt} = \frac{1}{i}[\hat{O},\hat{K}] ~~~~ \leftrightharpoons ~~~~ \dot{O} = \{O,H\}.
\end{equation}
The KvNS approach fully recovers classical dynamics.

If instead we once again begin with the Dirac-von Neumann axioms, Ehrenfest theorems (eqs. \ref{eq: ehrenfest_thm_1}, \ref{eq: ehrenfest_thm_2}), and instead use the canonical commutator,

\begin{equation}\label{eq: canonical_commutator}
    [\hat{x},\hat{p}]=i\hbar,
\end{equation}
we will recover quantum dynamics as the Schr{\"o}dinger equation. Starting with noncommuting position and momentum, we recover:
\begin{equation}
    [\hat{x},\hat{\Omega}(\hat{x},\hat{p})] = \frac{i\hat{p}}{m} ~~~~\Rightarrow ~~~~ [\hat{x},\hat{p}]\frac{\partial \hat{\Omega}}{\partial p} = \frac{i\hat{p}}{m} ~~~~\Rightarrow ~~~~ \hat{\Omega} = \frac{\hat{p}^2}{2m} + C,
\end{equation}
\begin{equation}
    [\hat{p},\hat{\Omega}(\hat{x},\hat{p})] = -iU'(\hat{x}) ~~~~\Rightarrow ~~~~ [\hat{p},\hat{x}]\frac{\partial \hat{\Omega}}{\partial x} = -i\frac{\partial U}{\partial x} ~~~~\Rightarrow ~~~~ \hat{\Omega} = U(\hat{x}) + C.
\end{equation}
This is a derivation of the well known Hamiltonian operator (relabeled as $\hat{H}$):

\begin{equation}
    \hat{H} = \frac{\hat{p}^2}{2m} + U(\hat{x}).
\end{equation}
Plugging into Stone's theorem gives us the Schr{\"o}dinger equation:
\begin{equation}
    \ket{\dot{\psi}} = -i\hat{H}\ket{\psi}.
\end{equation}
For a more detailed derivation of the above, see \citet{ODM} and review \citet{PiaseckiThesis}.

An important consequence of the summarized derivations above is that the difference between classical and quantum mechanics lies in the commutation of position and momentum. Using the ODM approach, we now are able to construct a unified classical-quantum algebra, which leads directly to a natural derivation of the Wigner quasiprobability distribution \citep{ODM, wignerphasespace, wave_operator}. To do this, we begin with a unified quantum-classical commutator of the form \citep{ODM,wignerphasespace},

\begin{equation}
    [\boldsymbol{\hat{x}},\boldsymbol{\hat{p}}]=i\hbar\kappa,
\end{equation}
and follow the same steps as above. In the limit of $\kappa$ goes to zero, we recover classical mechanics, and as $\kappa$ goes to one, we recover quantum mechanics. Notice we bolded the above operators. This is because we also must take into account the Koopman algebra (eq. \ref{eq: koopman_algebra}), which is usually not considered in the quantum case. The quantum operators (bolded) are constructed out of the classical operators (unbolded) in the following manner:

\begin{equation}
    \boldsymbol{\hat{x}}=\hat{x} -\frac{\hbar\kappa}{2}\hat{\lambda}_p,
\end{equation}
\begin{equation}
    \boldsymbol{\hat{p}} = \hat{p}+\frac{\hbar \kappa}{2}\hat{\lambda}_x,
\end{equation}
and we retain classical commutation relations. Starting with the Ehrenfest theorems, we are led to 

\begin{equation}
    [\boldsymbol{\hat{x}},\hat{\Omega}(\boldsymbol{\hat{x}},\boldsymbol{\hat{p}}, \hat{\lambda}_x,\hat{\lambda}_p)] = \frac{i\boldsymbol{\hat{p}}}{m} ~~~~\Rightarrow ~~~~ [\boldsymbol{\hat{x}},\hat{\lambda}_x]\frac{\partial \hat{\Omega}}{\partial \lambda_x}+[\boldsymbol{\hat{x}},\boldsymbol{\hat{p}}]\frac{\partial \hat{\Omega}}{\partial \boldsymbol{{p}}}=\frac{i\boldsymbol{\hat{p}}}{m}  ~~~~\Rightarrow ~~~~ \kappa m \frac{\partial \hat{\Omega}}{\partial \boldsymbol{p}}+\frac{m}{\hbar}\frac{\partial \hat{\Omega}}{\partial \lambda_x} = \boldsymbol{\hat{p}},
\end{equation}
\begin{equation}
    [\boldsymbol{\hat{p}},\hat{\Omega}(\boldsymbol{\hat{x}},\boldsymbol{\hat{p}}, \hat{\lambda}_x,\hat{\lambda}_p))] = -iU'(\boldsymbol{\hat{x}}) ~~~~\Rightarrow ~~~~ [\boldsymbol{\hat{p}},\hat{\lambda}_p]\frac{\partial \hat{\Omega}}{\partial \lambda_p} + [\boldsymbol{\hat{p}},\boldsymbol{\hat{x}}]\frac{\partial \hat{\Omega}}{\partial \boldsymbol{x}}= -iU'(\boldsymbol{\hat{x}}) ~~~~\Rightarrow ~~~~ \kappa \frac{\partial \hat{\Omega}}{\partial \boldsymbol{x}}-\frac{1}{\hbar}\frac{\partial \hat{\Omega}}{\partial \lambda_p} = U'(\boldsymbol{\hat{x}}).
\end{equation}
Through the method of characteristics, we recover the unified quantum-classical generator of motion:

\begin{equation}
    \hat{\Omega} = \frac{1}{\kappa}\Big(\frac{\boldsymbol{\hat{p}^2}}{2m} + U(\boldsymbol{\hat{x}})\Big)+F(\boldsymbol{\hat{x}} -\hbar\kappa\hat{\lambda}_p,\boldsymbol{\hat{p}}+\hbar\kappa\hat{\lambda}_x).
\end{equation}
The operator $\hat{F}$ at the tail end of the above generator commutes with all observables of the form $O(\boldsymbol{\hat{x},\boldsymbol{\hat{p}}})$, so does not impact experimental results. However, if we wish it to reduce to $\hbar\hat{K}$ in the limit $\kappa\rightarrow 0$ (classical limit), it must take on a very specific shape. With other important limits taken into consideration to maintain self-consistency (see \citealt{ODM}), it can be shown that $\hat{F}$ takes on a form such that the generator as a whole is (relabeling as $\hat{H}_{qc}$):

\begin{equation}\label{eq: nonrel_H_qc}
    \hat{H}_{qc} = \frac{\hbar}{m}\hat{p}\hat{\lambda}_x + \frac{1}{\kappa}U\Big(\hat{x}-\frac{\hbar\kappa}{2}\hat{\lambda}_p\Big) - \frac{1}{\kappa}U\Big(\hat{x}+\frac{\hbar\kappa}{2}\hat{\lambda}_p\Big).
\end{equation}
This gives us consistent limits and other aesthetically pleasing features \citep{ODM}. Out of this, we now derive the Wigner distribution.

Starting from unitary time evolution,
\begin{equation}
    \ket{\dot{\psi}} = -i\hat{H}_{qc}\ket{\psi},
\end{equation}
we sandwich both sides in the basis $\bra{x\lambda_p}$. Using the variable substitution $u = \hat{x}-\frac{\hbar\kappa}{2}\hat{\lambda}_p$ and $v=\hat{x}+\frac{\hbar\kappa}{2}\hat{\lambda}_p$, we get

\begin{equation}\label{eq: equal_time_split}
    \Big(i\hbar\kappa\frac{\partial}{\partial t} - \frac{(\hbar\kappa)^2}{2m}\Big(\frac{\partial^2}{\partial v^2}-\frac{\partial^2}{\partial u^2}\Big) -U(u) + U(v)\Big)\rho_\kappa = 0.
\end{equation}
Notice that $\rho_\kappa(u,v,t) \propto \braket{x\lambda_p|\psi(t)}$. Switching to an $xp$ representation we achieve a quantum amplitude on classical phase space,

\begin{equation}\label{eq: density_to_amplitude}
    \braket{xp|\psi} = \sqrt{\frac{\hbar\kappa}{2\pi}}\int d\lambda_p~\rho_\kappa(\hat{x}-\frac{\hbar\kappa}{2}\hat{\lambda}_p,\hat{x}+\frac{\hbar\kappa}{2}\hat{\lambda}_p)e^{ip\lambda_p}.
\end{equation}
In the classical limit, the phase space wavefunction reduces to the classical KvNS wavefunction. In other words, the Wigner distribution corresponds to a \textit{probability amplitude} of a quantum particle to be found at a point in classical phase space, instead of a quasiprobability distribution \citep{wignerphasespace}. Since wavefunction amplitudes can be negative, the Wigner distribution's lack of positive definiteness ceases to be a conundrum \citep{wignerphasespace, wave_operator}.

The Moyal's equation of motion can further be derived from eq. \ref{eq: nonrel_H_qc}. Recall that the Moyal bracket is defined with directional derivatives as

\begin{equation}
    \{\{f,g\}\} = \frac{2}{\hbar\kappa}f(x,p)\sin\Big(\frac{\hbar\kappa}{2}\overset{\leftarrow}{\frac{\partial}{\partial x}}\overset{\rightarrow}{\frac{\partial}{\partial p}}-\frac{\hbar\kappa}{2}\overset{\leftarrow}{\frac{\partial}{\partial p}}\overset{\rightarrow}{\frac{\partial}{\partial x}}\Big)g(x,p).
\end{equation}
Keeping in mind the identity $\exp(a\vec{\partial}_y)f(y) = f(y)\exp(a\overset{\leftarrow}{\partial}_y) = f(y+a)$, we can write the Moyal bracket as \citep{wignerphasespace}

\begin{equation}
     \{\{f,g\}\} = \frac{1}{i\hbar\kappa}\bra{xp}\Big(f\big(\hat{x} -\frac{\hbar\kappa}{2}\hat{\lambda}_p,\hat{p}+\frac{\hbar \kappa}{2}\hat{\lambda}_x)-f(\hat{x} +\frac{\hbar\kappa}{2}\hat{\lambda}_p,\hat{p}-\frac{\hbar \kappa}{2}\hat{\lambda}_x)\Big)\ket{g}.
\end{equation}
Here, we clearly see the introduced operators $\hat{\lambda}_x$ and $\hat{\lambda}_p$ are the well-known Bopp operators \citep{bopp}. We can rewrite eq. \ref{eq: nonrel_H_qc} with Stone's theorem as the well-known expressions,
\begin{equation}
    \frac{\partial W}{\partial t}=\{\{H,W\}\}, ~~H = \frac{p^2}{2m}+U(x).
\end{equation}
We keep in mind that one of the attractive features of the Moyal bracket is their nice classical limit, leading to the Poisson bracket \citep{wignerphasespace}. ODM provides an elegant, unified, internally consistent picture for the non-relativistic Schr{\"o}dinger field.


%

\section{Relativistic Koopman Treatment of Wigner Operator with Abelian Gauge Coupling}

The Schr{\"o}dinger field is non-relativistic, so it is now important to generalize the previous results to a model consistent with the theory of relativity. One of the issues with a relativistic view of ODM is that relativity treats time and space on equal footing, whereas in the non-relativistic quantum case we treat time as a parameter and space as operatorial. The mixing of space and time creates two possible avenues for the ODM treatment. One, consistent with all previously successful relativistic quantum field theories, we demote space to being a parameter. In the second approach, we define a type of time operator, and discuss limitations. Whether one takes a field-theoretic approach or a four-operator approach to the ODM treatment, one derives the same conclusions, as we will see. 

We develop the relativistic cases for applications to QCD. PDFs, GPDs, and TMDs are derived from the Wigner distribution (eq. \ref{eq: wigner_function}) for studying quarks, antiquarks, and gluons \citep{Quark_Distribution_9,QCD_quark_gluon_1986_HE,quark_dist_book,Quark_Distribution_11,Wigner_Operator_extra_reference,GPDs, Quark_Distributions_2,Quark_Distributions_3, Quark_Distributions_1, Quark_Distributions_4,quark_wigner_2017,gluon_wigner_1,quark_wigner_2018,wigner_phd_thesis,Quark_Distribution_5,Quark_Distribution_7,Quark_Distribution_6}. Of central importance to our analyses of quarks and gluons is the relativistic Wigner operator \citep{Quark_Distribution_9,quark_dist_book,Quark_Distribution_11,Wigner_Operator_extra_reference,GPDs, Quark_Distributions_2,Quark_Distributions_3, Quark_Distributions_1, Quark_Distributions_4,wigner_phd_thesis,Quark_Distribution_5,Quark_Distribution_7,Quark_Distribution_6,covariant_wigner_QCD_2021}, which is generally defined as


\begin{equation}\label{eq: relativistic_wigner_function}
    \hat{W}({X^\mu},P^\nu) = \int d^4\Theta~\Psi\Big({X^\mu}+\frac{{\Theta^\mu}}{2}\Big)\Bar\Psi\Big({X^\mu}-\frac{{\Theta^\mu}}{2}\Big)e^{-iP^\nu\Theta_\nu},
\end{equation}
where ${X^\mu}$ is the four-position in spacetime, $P^\nu$ is the four-momentum, and ${\Theta^\mu}$ is a spacetime separation. Since quarks are fermions, the above $\Psi$ represent Dirac spinors with color degrees of freedom. Importantly, the above Wigner operator contains the time variable \citep{Wigner_Operator_extra_reference,GPDs, Quark_Distributions_2,Quark_Distribution_5,Quark_Distribution_7}, due to the equal treatment of time and space. The advantage of a fully covariant Wigner operator (as opposed to equal time Wigner operator) is the potential use of interpolating forms of dynamics \citep{rel_quantum_invariance_Ji_2025}. The operator is Lorentz covariant \citep{Quark_Distribution_5,Quark_Distribution_6} and defined on the eight dimensional phase space $(X^\mu,P^\nu)$.

We begin with the Abelian case, which will serve as the foundation for the non-Abelian generalization discussed in Section \ref{nonabelian_section}. We simplify the notation in one important way from the previous section, and that is instead of having an interpolating parameter $\kappa$ in what follows, we will either take the limit of $\hbar \rightarrow 0$ for the classical case or $\hbar \rightarrow 1$ for the quantum case (keeping in mind natural units). Natural units are assumed, but will retain $\hbar$ in the equations where relevant. We assume metric signature $\eta^{\mu\nu}=diag[1,-1,-1,-1]$ all throughout this work.


\subsection{Four-Operator First Quantized Approach}\label{First_Quant_section}

In this section we generalize the analysis of the nonrelativistic case by introducing operators of the form $X^\mu,~P^\mu,~\Lambda^\mu, ~\Theta^\mu$, etc., which we refer to as relativistic four-operators. This construction implicitly assumes a Hermitian time component of $X^\mu$, which of course by Pauli's theorem, has debatable mathematical rigor. However, we take the four-operator approach as a heuristic device that motivates the field-theoretic formulation discussed in Section \ref{field_theoretic_1}, where space and time are equally demoted to parameters. For the moment we proceed under the assumption that these four-operators exist, and return later to the question of the Hermitian time operator.

The inhomogeneous Lorentz (Poincar\'e) group governs the symmetries of spacetime and provides the foundation for most field-theoretic constructions. Starting from the Poincar\'e transformation,
\begin{equation}
    x^\mu = \Lambda^\mu_\nu x^\nu + a^\mu,
\end{equation}
one can derive the classical (pre-quantized) Poincar\'e commutation relations,

\begin{equation}\label{eq: first_commutator}
    [J^j,J^k]=-i\epsilon^{jkl}J^l,
\end{equation}
\begin{equation}
    [K^j,K^k]=-i\epsilon^{jkl}J^l,
\end{equation}
\begin{equation}
    [J^j,K^k]=i\epsilon^{jkl}K^l,
\end{equation}
\begin{equation}
    [P^\mu,P^\nu]=0,
\end{equation}
\begin{equation}
    [K^j,P^0]=-iP^j,
\end{equation}
\begin{equation}
    [J^j,P^0]=0,
\end{equation}
\begin{equation}
    [K^k,P^j]=-i\delta_{jk}P^0,
\end{equation}
\begin{equation}\label{eq: last_commutator}
    [J^j,P^k]=i\epsilon^{jkl}P^l.
\end{equation}
These commutation relations come about before any assumptions about quantum or classical natures have been made. They are universal commutation relations that arise strictly from the group properties of the spacetime transformation \citep{dr_ji_book}. To introduce quantum versus classical scenarios, one must introduce \textit{an additional commutator} beyond the above group theoretic ones \citep{dr_ji_book}. The quantum case arises from the additional commutator,

\begin{equation}\label{eq: four_commutator}
    [X^\mu,P_\nu] = -i\hbar\delta^\mu_\nu.
\end{equation}
Or, equivalently, introducing to the above commutation relations \citep{dr_ji_book},

\begin{equation}
    \bra{X^\nu}P_\mu = i\hbar\delta^\nu_\mu\frac{\partial}{\partial X^\nu}\bra{X^\nu},
\end{equation}
where the repeated index does not imply summation.
One can then take the same approach as in the nonrelativistic ODM case, with a universal commutator of $[X^\mu,P_\nu] = -i\hbar\delta^\mu_\nu$. In the limit $\hbar \rightarrow 0$, we recover classical dynamics, and in the limit $\hbar \rightarrow 1$, one recovers relativistic quantum theory from a parallel series of steps. For the classical case \citep[]{ODM,OpenDirac}, we maintain, 

\begin{equation}\label{eq: classical_commutation_of_xp}
    [X^\mu,P_\nu]=0,
\end{equation}
with the relativistic Koopman algebra 
\begin{equation}\label{eq: classical_rel_commutators}
    [X^\mu,\Lambda_\nu]= -i\delta^\mu_\nu, ~[P_\mu,\Theta^\nu] = -i\delta^\nu_\mu.
\end{equation}
Just as before, but now for the relativistic case, we will develop the  Wigner distribution. 

The concept of classical spin remains an active area of discussion (\citealt{spinorial_1,Ohanian_1986,spinorial_2,spinorial_4,spinorial_5,classical_spin_theory_2014,Hestenes_book_spinors,barandes_qft_1,barandes_qft_2}, etc.). In the unified framework presented here, when discussing fermions, we introduce a spinor (using the notation of \citealt{relativistic_density}) that serves as a Koopman-type amplitude in the classical limit: 

\begin{equation}\label{eq: Dirac_spinor}
    \ket{\Psi} = \begin{bmatrix}
    \ket{\Psi, 1} \\
    \ket{\Psi, 2} \\
    \ket{\Psi, 3} \\
    \ket{\Psi, 4} 
    \end{bmatrix}.
\end{equation}
The use of spinors in a classical setting is well established.
Historically, spinors first were developed for classical rigid body dynamics where they provide an efficient description of orientation and rotational motion \citep{Clifford_algebra_book_w_spinors}. 
Within Clifford algebra frameworks developed by \citet{spinorial_1,spinorial_3}, \citet{spinorial_2}, \citet{spinorial_4}, and others, classical spinors arise as geometric objects encoding orientation and momentum in a unified manner on spacetime. In this sense, they furnish a natural language for relativistic classical dynamics.

The distinction between classical and quantum spin then lies not in the mathematical structure of the spinor itself, but in the nature of the associated observables.
Classical spin is a continuum, whereas quantum spin is restricted to discrete, quantized values \citep{spinorial_2, barandes_qft_1, barandes_qft_2}. In this sense, classical spin may be viewed as the large occupancy limit of quantum spin states \citep{Ohanian_1986,barandes_qft_2}, consistent with Bohr’s correspondence principle. For example, the spin of a classical electromagnetic wave can be seen as the high occupancy limit of photon spin \citep{Ohanian_1986}.
This perspective aligns naturally with the Koopman interpretation advanced here.


Since relativity blends space and time together, time must no longer hold an esteemed position over spatial coordinates. The way we presented Stone's theorem  and the Ehrenfest theorems in the nonrelativistic section suggests time evolution has a unique role over the role $\hat{x}$ takes. To remedy this, instead of using an Ehrenfest-like theorem, we replace this ingredient with an equivalent ingredient of infinitesimal generators of motion \citep{Forms_of_Dynamics} of the form,

\begin{equation}
    \zeta = \zeta_0 + [\zeta_0,F],
\end{equation}
where $F$ is an infinitesimal generator of motion and $\zeta_0$ is a dynamical variable of interest. This is the exact formula that \citet{Forms_of_Dynamics} used when developing the forms of relativistic dynamics (point, front, and instant) to get the previous Poincar\'e commutation relations (eqs. \ref{eq: first_commutator}-\ref{eq: last_commutator}). It is also equivalent to a relativistic Heisenberg equation of motion (see, for example, \citealt{heisenberg_eqs_motion_1}). We postulate a generator of motion $D(X^\mu,P_\nu)$ that obeys \citep{OpenDirac},

\begin{equation}\label{eq: D_psi}
    D(X^\mu,P_\nu)\ket{\Psi}=0,
\end{equation}
where $\ket{\Psi}$ is our spinorial ket. 
This postulate is equivalent to postulating the form of $\hat{H}$ as $H(\hat{x},\hat{p})$ in the nonrelativistic case.

In place of the Ehrenfest theorems, we deduce \textit{by ansatz} the evolution of $X^\mu$ is provided by

\begin{equation}\label{eq: replacement_of_first_ehrenfest}
    [X^\mu,D(X^\nu,P_\sigma)] = i\gamma^\mu,
\end{equation}
and the evolution of $P^\mu$,

\begin{equation}\label{eq: replacement_of_second_ehrenfest}
    [P^\mu,D(X^\nu,P_\sigma)] = ie\gamma^\iota\partial^\mu A_\iota.
\end{equation}
This replaces the use of the Ehrenfest-like theorems postulated by \citet{relODM} in their derivation of the Dirac and Spohn equations. The above expressions are equivalent to the Heisenberg equations of motion for $dX^\mu/ds$ and $dP^\mu/ds$ (eq. 22 of \citealt{relODM}). 

If we assume the four-canonical commutator (eq. \ref{eq: four_commutator}),

\begin{equation*}
    [X^\mu,P_\nu]\frac{\partial D}{\partial P_\nu} = i\gamma^\mu ~~\Rightarrow~~ D(X^\mu,P_\nu)=-\gamma^\nu P_\nu + C(X^\mu),
\end{equation*}
\begin{equation*}
    [P^\mu,X_\nu]\frac{\partial D}{\partial X_\nu} = ie\gamma^\iota\partial^\mu A_\iota ~~\Rightarrow~~ D(X^\mu,P_\nu)=e\gamma^\iota A_\iota + C(P_\nu).
\end{equation*}
Together, our covariant Dirac generator is

\begin{equation}\label{eq: rel_Dirac_generator}
    D(X^\mu,P_\nu)=-\gamma^\nu P_\nu +e\gamma^\iota A_\iota + C.
\end{equation}
As discussed in \citet{relODM} and the appendix of \citet{OpenDirac}, considerations of the classical limit require the remaining constant equal

\begin{equation}\label{eq: rel_Driac_generator_2}
    C = m.
\end{equation}
Combining eqs. \ref{eq: D_psi}, \ref{eq: rel_Dirac_generator}, and \ref{eq: rel_Driac_generator_2}, we have derived the full covariant Dirac equation. 

Now, taking the Koopman approach, we begin with eqs. \ref{eq: replacement_of_first_ehrenfest} and \ref{eq: replacement_of_second_ehrenfest} once more. Since now $X^\mu$ and $P^\nu$ commute, we work with a Hilbert phase space \citep{ODM,wignerphasespace,OpenDirac,relODM}. If we assumed eq. \ref{eq: classical_commutation_of_xp} is the only classical commutator, we would run into the same contradiction as the nonrelativistic classical case (eqs. \ref{eq: contra_1}, \ref{eq: contra_2}). This necessitates the relativistic Koopman algebra \citep{ODM}. Following a parallel series of steps,

\begin{equation}
    D(X^\mu, P^\nu, \Lambda_\rho,\Theta_\sigma)\ket{\Psi}=0,
\end{equation}
\begin{equation}
    [X^\mu,\Lambda_\nu]\frac{\partial D}{\partial \Lambda_\nu} = i\gamma^\mu ~~\Rightarrow~~ D(X^\mu, P^\nu, \Lambda_\rho,\Theta_\sigma)=-\gamma^\nu \Lambda_\nu + C(X^\mu, P^\nu,\Theta_\sigma),
\end{equation}
\begin{equation}
    [P^\mu,\Theta_\nu]\frac{\partial D}{\partial \Theta_\nu} = ie\gamma^\iota\partial^\mu A_\iota ~~\Rightarrow~~ D(X^\mu, P^\nu, \Lambda_\rho,\Theta_\sigma)=-e\gamma^\iota (\partial^\sigma A_\iota)\Theta_\sigma + C(X^\mu, P^\nu, \Lambda_\rho).
\end{equation}
Therefore, together, the covariant Koopman generator (relabeled $K$)

\begin{equation}\label{eq: rel_koopman_generator}
    K(X^\mu, P^\nu, \Lambda_\rho,\Theta_\sigma) = -\gamma^\nu \Lambda_\nu - e\gamma^\iota (\partial^\sigma A_\iota)\Theta_\sigma + C(X^\mu, P^\nu).
\end{equation}
Eq. \ref{eq: rel_koopman_generator} represents the classical, relativistic Koopman theory. 

Since $X^\mu$ and $P^\nu$ commute in the classical limit, we have an eigenbasis $\bra{X^\mu P^\nu}$. Sandwiching from the left:

\begin{equation}\label{eq: kvn_rel_sandwich}
    \bra{X^\mu P^\nu}K\ket{\Psi}= \Big(-i\gamma^\nu \frac{\partial}{\partial X_\nu} - ie\gamma^\iota (\partial^\sigma A_\iota)\frac{\partial}{\partial P_\sigma} + C(X^\mu, P^\nu)\Big)\braket{X^\mu P^\nu|\Psi} = 0.
\end{equation}
The above is the phase space generalization of the Dirac operator with Koopman spinor. 
For the classical case, we set the constant of integration to zero. 

We must ensure the gauge invariance of the Koopman formulation. 
Note that since the momentum $P_\mu$ appearing in the Koopman wavefunction is interpreted as the \emph{kinetic} momentum (see also discussion in \citealt{QCD_quark_gluon_1986_HE,Quark_Distribution_11}), one expands the first derivative in eq. \ref{eq: kvn_rel_sandwich} by the chain rule. Since the kinetic momentum acquires a position dependence, we have,

\begin{equation}
    P_\mu^{kin}(X)=P_\mu^{can}-eA_\mu(X),
\end{equation}
treating the canonical momentum as position independent (as a canonical coordinate). This induces,
\begin{equation}
    \frac{\partial}{\partial X_\nu}\rightarrow \frac{\partial}{\partial X_\nu}-e(\partial_\nu A_\alpha)\frac{\partial}{\partial P^\alpha}.
\end{equation}
This provides justification for generalizing the Koopman minimal coupling principle explored in  \citet{mincoupling_KvN}:

\begin{equation}
\Lambda_\mu' = \Lambda_\mu - e (\partial_\mu A_\alpha)\frac{\partial}{\partial P^\alpha}.
\end{equation}
This ensures the Koopman expression is gauge invariant (see for comparison \citealt{rel_kinetic_theory_2013}).

With this structure in place, we make use of Clifford algebra methods to project onto the scalar sector and obtain the classical phase-space dynamics. Defining the spacetime momentum $P=P_\nu\gamma^\nu$ in  basis $\{\gamma^\nu\}$ and multiplying eq. \ref{eq: kvn_rel_sandwich} from the right, taking grade-0 elements (described in \citealt{CAS_doran_book,CAS_Spacetime_EM,Hestenes_book_spinors}), one recovers the typical relativistic Liouville flow \citep{rel_kinetic_theory_2013} in the form of
\begin{equation}\label{eq: Vlasov_flow}
     \Big(iP^\nu \frac{\partial}{\partial X_\nu} + ieP^\iota F^\sigma_\iota\frac{\partial}{\partial P_\sigma}\Big)\braket{X^\mu P^\nu|\Psi} = 0.
\end{equation}
Using the relativistic Born Rule $\rho(X^\mu,P^\nu) = \braket{\Psi|X^\mu P^\nu}\braket{X^\mu P^\nu|\Psi}$ to derive

\begin{equation}\label{eq: EVM_eq}
    \Big(P^\nu \frac{\partial}{\partial X_\nu} + eP^\iota F^\sigma_\iota\frac{\partial}{\partial P_\sigma}\Big)\rho(X^\mu,P^\nu) = 0.
\end{equation}
This is the well-known Vlasov-Maxwell equation for non-colliding point particles in flat spacetime with Abelian electromagnetic interaction \citep{VM_5,VM_3,VM_2,VM_1, VM_4}. 

Here, $\rho(X^\mu,P^\nu)=\Psi^\dagger \Psi$ represents classical particle number density on relativistic phase
space. 
As argued elsewhere, the canonical commutator encodes information about wavenature, and the commuting position and momentum observables leads to point particle behavior \citep{DvN_EM}. The Vlasov-Maxwell equation is an aesthetically elegant equation to derive in the classical limit, as it encapsulates the characteristics of relativistic classical point particle dynamics. As we will see, in the non-Abelian case, a similar Vlasov equation will be the correct classical limit of QCD in the form of a relativistic quark-gluon plasma, i.e. colored point particles obeying classical relativistic dynamics \citep{Wong_eq_1970,Wong_eq_1989,Wong_eq_1994, Wong_eq_1998, Wong_eq_1999_a, Wong_eq_2002,Wong_eq_2002_2,Wong_eq_2002_3, Wong_eq_2003,Wong_eq_2003_2,Wong_eq_2009,Wong_eq_2013}.

Next, we want a full picture of unified dynamics, just like with the nonrelativistic case. In the quantum limit, we want to derive the Dirac generator (eqs. \ref{eq: rel_Dirac_generator}, \ref{eq: rel_Driac_generator_2}) and in the classical, point particle limit we want the relativistic Koopman generator (eq. \ref{eq: rel_koopman_generator}). The quantum-classical four-canonical commutator is \citep{ODM}

\begin{equation}
[\boldsymbol{X}^\mu,\boldsymbol{P}_\nu]=-i\hbar\delta^\mu_\nu.
\end{equation}
As before, quantum operators (bolded) are built out of classical ones (unbolded):

\begin{equation}
    \boldsymbol{X}^\mu=X^\mu-\frac{\hbar}{2}\Theta^\mu,
\end{equation}
\begin{equation}
    \boldsymbol{P}_\nu = P_\nu + \frac{\hbar}{2}\Lambda_\nu.
\end{equation}
Classical commutators obey eq. \ref{eq: classical_rel_commutators}. Starting with the commutator relationships eqs. \ref{eq: replacement_of_first_ehrenfest} and \ref{eq: replacement_of_second_ehrenfest}, we are led to

\begin{equation}
    [\boldsymbol{X}^\mu,D(\boldsymbol{X}^\nu, \boldsymbol{P}^\sigma,\Lambda_\rho,\Theta_\iota)]=i\gamma^\mu ~~\Rightarrow~~[\boldsymbol{X}^\mu,\boldsymbol{P}^\sigma]\frac{\partial D}{\partial \boldsymbol{P}_\sigma}+[\boldsymbol{X}^\mu,\Lambda_\rho]\frac{\partial D}{\partial \Lambda_\rho}=i\gamma^\mu, 
\end{equation}
\begin{equation}
    [\boldsymbol{P}^\mu,D(\boldsymbol{X}^\nu,\boldsymbol{P}^\sigma,\Lambda_\rho,\Theta_\iota)] = ie\gamma^\iota\boldsymbol{\partial}^\mu A_\iota ~~\Rightarrow~~[\boldsymbol{P}^\mu,\boldsymbol{X}^\nu]\frac{\partial D}{\partial\boldsymbol{X}_\nu}+[\boldsymbol{P}^\mu,\Theta_\iota]\frac{\partial D}{\partial \Theta_\iota} = ie\gamma^\iota\boldsymbol{\partial}^\mu A_\iota.
\end{equation}
This results in
\begin{equation}
    -i\hbar\eta^{\sigma\mu}\frac{\partial D}{\partial \boldsymbol{P}_\sigma}+i\frac{\partial D}{\partial \Lambda^\mu}=i\gamma^\mu,
\end{equation}
\begin{equation}
    i\hbar\eta^{\mu\nu}\frac{\partial D}{\partial\boldsymbol{X}_\nu} + i\frac{\partial D}{\partial \Theta^\mu}=ie\gamma^\iota\boldsymbol{\partial}^\mu A_\iota.
\end{equation}
This has the general solution:

\begin{equation}
    D(\boldsymbol{X}^\nu, \boldsymbol{P}^\sigma,\Lambda_\rho,\Theta_\iota) = \frac{1}{\hbar}\Big(-\gamma^\mu \boldsymbol{P}_\mu + e\gamma^\iota A_\iota(\boldsymbol{X})
    +m\Big)+F(\boldsymbol{P}_\mu -\hbar \Lambda_\mu, \boldsymbol{X}_\nu +\hbar\Theta_\nu).
\end{equation}
Just like in the nonrelativistic case, the operator $F$ commutes with all observables of the form $O(\boldsymbol{X}^\mu,\boldsymbol{P}_\nu)$, so it does not modify any quantum experiment. In the classical-quantum generator above, we also included the constant $m\gamma^0$ so that we recover the Dirac generator in the limit of $\hbar \rightarrow 1$. 

Our task now is to find the shape of the relativistic $F$. \citet{ODM} derives the form of the nonrelativistic $F$ from various limit considerations (getting the complete generator, eq. \ref{eq: nonrel_H_qc}). For unified dynamics, we foremost want

\begin{equation}
    \lim_{\hbar\rightarrow0}D(\boldsymbol{X}^\nu, \boldsymbol{P}^\sigma,\Lambda_\rho,\Theta_\iota) = K(X^\mu, P^\nu, \Lambda_\rho,\Theta_\sigma).
\end{equation}
Let $\alpha_\mu=\boldsymbol{P}_\mu -\hbar \Lambda_\mu$ and $\beta_\nu = \boldsymbol{X}_\nu +\hbar\Theta_\nu$. We can then state

\begin{equation}
    \frac{\partial K}{\partial \Lambda_\mu}=\lim_{\hbar\rightarrow0}\frac{\partial D}{\partial \Lambda_\mu} ~~\Rightarrow~~\gamma^\mu = \lim_{\hbar\rightarrow 0}\frac{1}{\hbar}\Big(-\gamma^\epsilon\frac{\partial \boldsymbol{P}_\epsilon}{\partial \Lambda_\mu}\Big)+\frac{\partial F}{\partial \alpha_\sigma}\Big(\frac{\partial \boldsymbol{P}_\sigma}{\partial \Lambda_\mu}-\hbar\Big),
\end{equation}
\begin{equation}
    \frac{\partial K}{\partial \Theta^\iota}=\lim_{\hbar\rightarrow0}\frac{\partial D}{\partial \Theta^\iota} ~~\Rightarrow~~e\gamma^\sigma(\partial^\iota A_\sigma) = \lim_{\hbar\rightarrow 0}\frac{1}{\hbar}\Big(e\gamma^\eta\frac{\partial A_\eta}{\partial \boldsymbol{X}^\sigma}\frac{\partial \boldsymbol{X}^\sigma}{\partial \Theta^\iota}\Big)+\frac{\partial F}{\partial \beta_\sigma}\Big(\frac{\partial \boldsymbol{X}^\sigma}{\partial \Theta^\iota}+\hbar\Big).
\end{equation}
This suggests that in the limit of $\hbar \rightarrow 0$, the form of $F$ is

\begin{equation}
    F(P_\nu,X_\mu) = \frac{\gamma^\nu}{\hbar}P_\nu - \frac{e\gamma^\sigma}{\hbar}A_\sigma(X_\mu) + O(1).
\end{equation}
Taking the leading terms as the definition of the shape of $F$ and re-expressing it in terms of classical operators, simplifying, we get:
\begin{equation}
    D= -\gamma^\nu\Lambda_\nu - \frac{m}{\hbar} + \frac{e\gamma^\epsilon}{\hbar}\Big(A_\epsilon(X^\mu-\hbar\Theta^\mu/2)-A_\epsilon(X^\mu+\hbar\Theta^\mu/2)\Big).
\end{equation}
We make one more modification, namely, the mass term in the above expression will be rather awkward to handle. To retain it in the quantum limit (eqs. \ref{eq: rel_Dirac_generator}, \ref{eq: rel_Driac_generator_2}), we must have the unobservable $F$ cancel this term in the classical limit. An experimentally undetectable modification would be

\begin{equation}
    F\rightarrow F+m/\hbar.
\end{equation}
This step is also important since, naively, the classical limit $\hbar \rightarrow 0$ for the term $m/\hbar$ will blow up to infinity. We expect any observable to commute with this expression, giving us the same conclusions as before.

We act from the left hand side with bra $\bra{X_\mu\Theta_\epsilon}$, and then rewrite $u^\mu = X^\mu - \hbar\Theta^\mu/2$ and $v^\mu = X^\mu +\hbar\Theta^\mu/2$ to get

$$\bra{X_\mu\Theta_\epsilon}D\ket{\Psi}= 0 ~~\Rightarrow~~$$
\begin{equation}\label{eq: u_v_exp}
\Big[i\hbar\gamma^\mu\Big(\frac{\partial}{\partial u^\mu}+\frac{\partial}{\partial v^\mu}\Big)+e\gamma^\epsilon\Big(A_\epsilon(u^\mu)-A_\epsilon(v^\mu)\Big)\Big]\braket{u^\mu v^\mu|\Psi}=0.
\end{equation}
The above expression is the exact same one provided by the relativistic density operator \citep{relativistic_density}. The relativistic von Neumann equation is given by \citep{relativistic_density},

\begin{equation}\label{eq: rel_density_operator}
    i\partial_0\bar{\rho}=[H_D,\bar{\rho}] ~~\leftrightharpoons~~ D\bar{\rho}-\bar{\rho}^\dagger D^\dagger=0,
\end{equation}
where $D= \gamma^\mu p_\mu - \gamma^\mu A_\mu -m$ and $\bar{\rho}=\hat{\rho}\gamma^0$ is the relativistic density operator, built out of spinors (eq. \ref{eq: Dirac_spinor}):

\begin{equation}\label{eq: dirac_density_eq_def}
    \hat{\rho}= \sum_{n=1}^{4} \ket{\Psi, n}\bra{\Psi, n}.
\end{equation}
Sandwiching eq. \ref{eq: rel_density_operator} on both sides with $\ket{u^\mu}, \ket{v^\nu}$, we get

\begin{equation}\label{eq: dirac-density}
    \Big[i\hbar\gamma^\mu\Big(\frac{\partial}{\partial u^\mu}+\frac{\partial}{\partial v^\mu}\Big)+ e\gamma^\epsilon\Big(A_\epsilon(u^\mu)-A_\epsilon(v^\mu)\Big)\Big]\bra{u^\mu}\hat{\rho}\gamma^0\ket{v^\mu}=0.
\end{equation}
Note that eqs. \ref{eq: rel_density_operator}-\ref{eq: dirac-density} make no assumption about a time operator, and are part of standard relativistic quantum analysis. 
We see by eqs. \ref{eq: u_v_exp} and \ref{eq: dirac-density} the quantities obey the same differential equations. Unlike the 1x1 case for nonrelativistic theory, however, with the relativistic case we deal with internal degrees of freedom and matrices. So simply comparing them is not enough. Worse, the Koopman spinor (in parallel to the Dirac spinor) is a 4x1 spinor of complex numbers, hence containing 8 real degrees of freedom. The coordinate density matrix in eq. \ref{eq: dirac-density} is a 4x4 complex matrix, which naively can have a maximum of 32 real degrees of freedom. Is it possible to map one into the other? 

There is a straightforward and natural way to relate the two quantities above, realizing the Dirac density matrix encodes redundant information. {
Since the Wigner operator is Dirac Hermitian, this reduces to 16 real degrees of freedom. Furthermore, the Wigner operator obeys the additional kinetic constraint in the collisionless limit \citep{QCD_quark_gluon_1986_HE,Quark_Distribution_11,wigner_phd_thesis,Quark_Distribution_7},

\begin{equation}\label{eq: kinetic_constraint}
(\gamma^\mu K_\mu -m)\hat W =0,
\end{equation}
where 
\begin{equation}
    K^\mu = \Pi^\mu +\frac{i\hbar}{2}\nabla^\mu,~~\Pi^\mu =p^\mu-\frac{\hbar}{2}j_1(\Delta)F^{\mu\nu}\partial_{p\nu},~~\nabla^\mu = \partial^\mu_x-j_0(\Delta)F^{\mu\nu}\partial_{p\nu},
\end{equation}
where $\Delta=\frac{\hbar}{2}\partial_\sigma^x\partial^\sigma_p$ and $j_0=\sin(x)/x$, $j_1=(\sin(x)-x\cos(x))/x^2$ are spherical Bessel functions.
The constraint further reduces the 16 to 8 real degrees of freedom \citep{Quark_Distribution_11}, the same number of degrees of freedom as a complex spinor. } Since both the interpolating spinorial object and $\hat W$ obey the same evolution equations (eqs. \ref{eq: u_v_exp}, \ref{eq: dirac-density}), this strongly suggests the existence of a correspondence between their independent degrees of freedom.

A rigorous realization of this correspondence can be obtained using the Clifford algebra structure underlying the Dirac theory. 
The Wigner operator can be expanded in the Dirac Clifford algebra basis:
\begin{equation}\label{eq: clifford_wigner}
   \hat  W=\mathcal{F}+\gamma^\mu V_\mu + \frac{1}{2}\sigma^{\mu\nu}S_{\mu\nu} + \gamma^5\gamma^\mu A_\mu+i\gamma^5\mathcal{P}
\end{equation}
where $\mathcal{F}, V_\mu, S_{\mu\nu},  A_\mu$ and $\mathcal{P}$ are real functions on relativistic phase space. 
Spinors are mathematically defined as the minimal left ideals of a Clifford algebra \citep{Lounesto_2001,Clifford_algebra_book_w_spinors}. To construct the minimal left ideal of the Dirac algebra we utilize an idempotent $\epsilon$, which is a constant projector (to be discussed further in the next section). Closely related, one can also define a spinor operator which is an element of the even Clifford subalgebra, denoted $Cl_{(1,3)}^+$ \citep{hestenes_dirac_1975,Lounesto_2001,Clifford_algebra_book_w_spinors}. Elements of the even subalgebra are also termed even multivectors \citep{CAS_doran_book,Hestenes_book_spinors}.

Due to the constraints on the Wigner operator, the 8 independent real degrees of freedom can be treated as $\mathcal{F}, \mathcal{P},$ and $S_{\mu\nu}$, with $V_\mu$ and $A_\mu$ as dependent variables \citep{Quark_Distribution_11}. Remarkably, degrees of freedom $\mathcal{F}, \mathcal{P},$ and $S_{\mu\nu}$ correspond to the even rank (or grade) expansion of the Wigner operator, denoted $\hat W_+$, and the constraints correspond to the odd grade, $\hat W_-$. \citet{hestenes_dirac_1975} demonstrates how the Dirac spinor originates from a spinor operator containing a scalar $\mathcal{F}$, pseudoscalar $\mathcal{P}$, and tensorial (bivector) $S_{\mu\nu}$ degrees of freedom, with the chief difference in that these degrees of freedom are a function of position or its Fourier dual momentum. The relationship between the even or odd rank (grades) can be demonstrated to be (in the limit of $\hbar \rightarrow 0$)

\begin{equation}
    \hat{W}_-=\frac{\gamma^\mu p_\mu}{m}\hat{W}_+,
\end{equation}
or,

\begin{equation}\label{eq: W_mapping}
    \hat{W}=\Big(1+\frac{\gamma^\mu p_\mu}{m}\Big)\hat{W}_+.
\end{equation}
The information in the Wigner operator is therefore in one-to-one correspondence to the even grade $\hat{W}_+$, which represents the information in a non-redundant way. In this context, the $\hat{W}_+$ corresponds to the matrix spinor operator \citep{Lounesto_2001}. To get a traditional spinor as an element of $\mathbb{C}^4$, we take the spinor operator and act on it from the right with the idempotent, $\Psi(x,p) = \hat{W}_+\epsilon$. These mappings are described in detail in Appendix II. We note that working with $\hat{W}_+$ rather than $\hat{W}$ directly provides the minimal representation with the correct spinorial degrees of freedom for Poincar\'e covariance as a minimal left ideal, as will be explored in an upcoming work. The mathematically well-defined spinor on phase space is therefore isomorophic to the Wigner operator.  

\begin{equation}\label{eq: ultimate_isomorphism}
    \hat{W}\cong \hat{W}_+
\cong \Psi(x,p).\end{equation}
A similar set of arguments exist for when the classical limit is not taken, but now with quantum corrections in terms of $\hbar$ to the spinor on phase space. 

{

%

In a similar fashion as to the nonrelativistic limit, we can then derive the Wigner matrix expression through a Koopman Fourier transform (to be discussed in the next section, eqs. \ref{eq: phase_space_Fourier}, \ref{eq: phase_space_Fourier_2}) of the Dirac density matrix on coordinate space:
\begin{equation}\label{eq: true_wigner}
    \hat{ W}(X,P)= \int d^4\Theta ~\bra{X^\mu - \hbar\Theta^\mu/2}\hat{\rho}\gamma^0\ket{X^\mu +  \hbar\Theta^\mu/2}e^{-i\Theta_\sigma P^\sigma}.
\end{equation}
The interpolating spinor on phase space is therefore obtained through the projection,
\begin{equation}
     \Psi(x,p) \cong \hat W,~~\Psi(x,p)=\hat W_+(x,p)\epsilon, ~~Cl^+_{(1,3)}(\mathbb{C})\epsilon\cong \mathbb{C}^4.
\end{equation}
We derive the Wigner operator, and demonstrate that it is isomorphic to the interpolating (quantum-classical) spinor on phase space, which maps to a Koopman spinor when $\hbar \rightarrow 0$. We note that in the limit of $\hbar \rightarrow 0$, eq. \ref{eq: kinetic_constraint} reduces to a classical on-shell mass constraint \citep{wigner_phd_thesis,Quark_Distribution_7}. Furthermore, as explored in \citet{wigner_phd_thesis} and \citet{Quark_Distribution_7}, the classical limit corresponds to the correct relativistic Liouville flow (also \citealt{QCD_quark_gluon_1986_HE,Wong_eq_1989,wigner_function_1994_OF,Quark_Distribution_11,wigner_classical_limit_2011}), so Koopman spinors will obey the Vlasov equation (eqs. \ref{eq: kvn_rel_sandwich}, \ref{eq: Vlasov_flow}).
\begin{equation}\label{eq: wigner_flow}
       \Big(P^\mu\partial_\mu +eP^\mu F_{\mu\nu}\frac{\partial}{\partial P_\nu}\Big)\hat W_{\hbar\rightarrow 0}=0.
\end{equation}
In the classical limit, $\hat{W}$ (and by extension, its isomorphic phase space spinor $\Psi$) encodes the correct Bargmann–Michel–Telegdi (BMT) classical spin dynamics \citep{wigner_phd_thesis,Quark_Distribution_7}. }

This sheds light how the Wigner operator, despite its matrix form, similarly acts as an interpolation between quantum and classical realms. In the relativistic limit, we still retain a similar interpretation as in the nonrelativistic limit, despite the multiple degrees of freedom. 
This correspondence shows us how a spinorial object relates to the Wigner operator. The above analysis would hold true once we consider non-Abelian fields (see \citealt{Quark_Distribution_11}).

We emphasize that there are two different Hilbert spaces involved in the above calculations. There is a Hilbert phase space (eq. \ref{eq: kvn_rel_sandwich}, left hand of eq. \ref{eq: true_wigner}) and regular quantum Hilbert space (where position and momentum do not commute, eq. \ref{eq: dirac-density}). The Wigner function is an interpolation between these two Hilbert spaces, and in the classical limit, transforms into a classical wavefunction under idempotent instead of a classical phase space distribution \citep{ODM,wignerphasespace}. Hence, the lack of positive definiteness for $\braket{X^\mu P^\nu|\Psi}$ is not an issue.

To finalize the derivation for the Abelian Wigner case, we need to impose gauge invariance \citep{QCD_quark_gluon_1986_HE,Quark_Distribution_11,GPDs,wigner_phd_thesis}. As it stands, eq. \ref{eq: true_wigner} is fully covariant, but as presented it is not yet gauge invariant. To do this is rather straightforward, by utilizing the identity \citep{wigner_phd_thesis}

\begin{equation}
    \Psi(X^\mu-\Theta^\mu/2)=\exp(-\frac{1}{2}\Theta_\nu\partial^\nu_X)\Psi(X^\mu).
\end{equation}
Replacing the above partial derivative with the covariant derivative automatically satisfies gauge-invariance, naturally including it, and can be used to derive the usual Wilson gauge-link \citep{wigner_phd_thesis}. The Koopman derivation of the Abelian Wigner operator is complete. 

A persistent challenge to the above four-operator approach is its reliance on a Hermitian time operator. According to the conventional reading of Pauli’s theorem, the existence of such an operator would require the Hamiltonian to have an unbounded spectrum, apparently ruling time operators out of standard quantum theory. This claim has been debated for nearly a century: while some view it as a fundamental obstacle, others have discussed weaknesses in its assumptions. For example, \citet{G_time_operator} argues that Pauli’s reasoning rests on unwarranted premises and therefore does not prohibit a self-adjoint time observable. Even so, a derivation framed within a field-theoretic approach is highly desired, where time and space are both demoted to parameters of the theory. 
Although Koopman field theoretic ideas have been explored \citep{YM,classical_field_koopman_EC_2011,ODM, kvn_field_theory}, the derivation of the Wigner function from this perspective is still left to be desired. In the next section, we take a step in this direction.

Although the above analysis utilizes a time operator, it is also possible to carry out an ODM treatment where one retains time as a parameter and space is an operator (\citealt{relODM}). The resulting structure is similar, but with the important drawback that the corresponding Wigner function is not derived, but assumed to get the correct classical predictions for particles. 
\footnote{After nearing completion of this work, we realized that a similar first quantized derivation of the Wigner function with time operator independently exists in the work of \citet{ODM_draft}, per private communication with Denys Bondar.}



\subsection{Field-Theoretic Second Quantized Approach}\label{field_theoretic_1}

A fully covariant Koopman field-theoretic formulation tailored to classical
relativistic point particles remains conceptually subtle. While relativistic
extensions of the KvNS framework have
been proposed, these typically either retain spatial coordinates as operators \citep{relODM,KvNPoincare}
or emphasize classical field theories (for example, scalar Klein-Gordon
systems) rather than genuine point particle phase space dynamics \citep{YM, classical_field_koopman_EC_2011, classical_field_koopman_EG_2011}. Moreover,
existing approaches do not explicitly construct a framework in which space
and time enter symmetrically as parameters in a manner compatible with the covariant
Wigner operator structures relevant for QCD.

In forthcoming work, we develop a relativistically covariant Koopman field
theory for classical point particles that makes these structural features
explicit. Our construction is inspired by the second quantization perspective
introduced by \citet{schonberg_1,schonberg_3,schonberg_2}, but is formulated to interface directly with the
Wigner function analysis central to QCD phenomenology. In the present review,
we outline the key structural elements necessary for this connection.

 Although the four-operator formulation remains debated, particularly in light
of Pauli's theorem and the difficulties associated with defining a
self-adjoint time operator, it furnishes a manifestly covariant framework and a few key insights
that proves structurally useful in constructing a fully field-theoretic
Koopman treatment. 

Rather than beginning with a four-operator construction, we take the opposite perspective to motivate the field-theoretic approach. It is already well-established that the Wigner operator obeys a kinematic constraint (eq. \ref{eq: kinetic_constraint}) and in the $\hbar \rightarrow 0$ limit, a Liouville flow (eq. \ref{eq: wigner_flow}). As discussed in the previous section, the Wigner operator admits a decomposition in the Dirac Clifford algebra (eq. \ref{eq: clifford_wigner}). Based on the independent degrees of freedom of the Wigner operator, one can map to the even subalgebra,  identifying the independent content with the spinor operator \citep{hestenes_dirac_1975,Lounesto_2001}. The properties of the Clifford algebra allow us to construct mathematically well-defined spinors from the spinor operator through the use of an idempotent  \citep{Bohm_KvN,Lounesto_2001,Clifford_algebra_book_w_spinors}. This algebraic structure therefore naturally allows the construction of a spinorial phase space object directly from the Wigner operator (see worked example in Appendix II). The resulting object retains the spinorial structure of the Dirac theory while obeying the classical Liouville dynamics in the $\hbar \rightarrow 0$ limit. We refer to such objects as \textit{Koopman spinors}. The higher
order $\hbar$ corrections to such a spinor correspond to the general Wigner operator
being mapped into a \textit{quantum spinor amplitude} on relativistic phase space (under idempotent).

To build an interpolating spinorial object, we first replace the nonrelativistic commutator structure and problematic time operator with a relativistically consistent set of Fourier transformations. The typical Fourier transformation we term the \textit{coordinate Fourier transform}, given that it transforms between the $\boldsymbol{x}$ and the inverse space of conjugated momentum $\boldsymbol{p}$:

\begin{equation}\label{eq: coordinate_fourier_transform}
    \psi(\boldsymbol{p}) = \int d\boldsymbol{x} ~e^{i\boldsymbol{x}\cdot\boldsymbol{p}/\hbar} \psi(\boldsymbol{x}),
\end{equation}
where the dot product implies the relativistic dot product. This is very familiar, but implies a wavenature between the conjugate variables $\boldsymbol{x}$ and $\boldsymbol{p}$, and in the nonrelativistic limit, implies the canonical commutator relationship (eq. \ref{eq: canonical_commutator}). In KvNS, we want to deconjugate position and momentum, so there is no wavenature between them and therefore a well-defined phase space. This implies a Fourier transformation that is phase space preserving, where classical $x$ and $p$ are conjugate to $\lambda_x$ and $\lambda_p$, respectively (\citealt{Bohm_KvN}):

\begin{equation}\label{eq: phase_space_Fourier}
    \psi(\lambda_x,p) = \int dx~d\lambda_p~e^{i\lambda_x \cdot x - \lambda_p \cdot p}\psi(x,\lambda_p),
\end{equation}
\begin{equation}\label{eq: phase_space_Fourier_2}
    \psi(x,\lambda_p) = \int d\lambda_x~dp~e^{-i\lambda_x \cdot x + \lambda_p \cdot p}\psi(\lambda_x,p).
\end{equation}
We consider a coordinate Fourier transform for quantum mechanics and a phase space preserving Fourier transform for Koopman classical mechanics. Any classical wavefunction must have its wavenature between conjugate quantities $x,\lambda_x$ and $p,\lambda_p$. The quantities $\lambda_x$ and $\lambda_p$ can then be further interpreted not just as Bopp operators \citep{ODM,wave_operator}, but as the inverse Fourier space of \textit{classical} $x$ and $p$ (Appendix I).

To construct spinors as mathematically well-defined minimal left ideals of the Clifford algebra, one introduces a primitive idempotent 
$\epsilon$ and multiplies algebra elements from the right, satisfying projector property $\epsilon^2 =\epsilon$. A representative choice of such an idempotent is \citep{Lounesto_2001,Clifford_algebra_book_w_spinors,CAS_idempotent_2015}

\begin{equation}\label{eq: second_idempotent}
    \epsilon = \frac{1}{4}(1+\gamma_0)(1+i\gamma_1\gamma_2).
\end{equation}
Note that different choices of idempotents will lead to isomorphic spinors. 
We choose for the idempotent to act on the even subalgebra $\hat W_+$ (which encodes the exact same independent degrees of freedom), in parallel to how the Dirac spinor is constructed in Clifford algebra frameworks as an even multivector under idempotent \citep{hestenes_dirac_1975,Lounesto_2001}.
The resulting mathematical object $\hat W_+ \epsilon$ therefore satisfies the criteria for a mathematically well-defined spinor, which is a function on relativistic phase space, and obeys classical Liouville flow when $\hbar \rightarrow 0$. We have identified a mathematically well-defined Koopman spinor, $\Psi(x,p) \in \mathbb{C}^4$, in close analogy to the nonrelativistic treatment \citep{wignerphasespace, wave_operator}. As noted in Appendix I, to derive a phase space from configuration space in Koopman theory, a doubling of variables is required. Recovering a phase space spinor from a coordinate density matrix, by removing redundant representation through use of an idempotent, is less strange in this light. Note the implicit presence of the Koopman phase preserving Fourier transform in the standard definition of the Wigner operator (eq. \ref{eq: relativistic_wigner_function}). This suggests that such a Koopman spinor is related to the waves in phase space described by eqs. \ref{eq: phase_space_Fourier}, \ref{eq: phase_space_Fourier_2}. 

The properties of the Koopman spinor construction will be explored further in the forthcoming paper, but, this initial look suggests that a purely field-theoretic approach to Koopman waves is possible and may shed light on the meaning of the Wigner operator. In the forthcoming work we will explore common QFT techniques for purely classical Koopman waves for point particles. 


The preceding construction suggests that the Wigner operator may be interpreted as a spinorial amplitude on relativistic phase space, interpolating between quantum and classical dynamics. This perspective naturally motivates a field-theoretic formulation of Koopman waves. We therefore conclude this section with a brief sketch of a second-quantized derivation of the Wigner operator.
To do this, we appeal directly to Heisenberg time evolution, 
    \begin{equation}\label{eq: Koopman_Heisenberg_equation_of_motion}
    \Psi'(t) = e^{it\Omega/\hbar} \Psi(0)e^{-it\Omega/\hbar}~~~\rightleftharpoons ~~~ i\hbar\partial_0\Psi= [\Psi,\Omega],
\end{equation}
where $\Omega$ is the generator of time evolution in the quantum case (as the Dirac Hamiltonian) or the classical case (as a relativistic Koopman-Liouville `Hamiltonian' or generator). Just like the previous analysis, we can also define a generator $\Omega(\hbar)$ that represents an interpolation of dynamics that encompasses both the classical ($\hbar  \rightarrow 0$) and quantum ($\hbar \rightarrow 1$). 
For Dirac spinors, we have the famous anticommutation relations,

\begin{equation}\label{eq: dirac_anticommutators}
    \{\Psi_\alpha(\boldsymbol{x}),\Psi_\beta^\dagger(\boldsymbol{x}')\}=\delta_{\alpha\beta}\delta^{(3)}(\boldsymbol{x}-\boldsymbol{x}'),
\end{equation}
\begin{equation}
    \{\Psi_\alpha^\dagger(\boldsymbol{x}),\Psi_\beta^\dagger(\boldsymbol{x}')\}=\{\Psi_\alpha(\boldsymbol{x}),\Psi_\beta(\boldsymbol{x}')\}=0.
\end{equation}
The Dirac Hamiltonian,

\begin{equation*}
    H_D= \int d^3\boldsymbol{p} ~\Psi^\dagger(\boldsymbol{p})\Big[\gamma_0\gamma_i (\boldsymbol{p}^i-eA^i(\boldsymbol{p}))+\gamma_0 mc^2+eA_0(\boldsymbol{p})\Big]\Psi(\boldsymbol{p}),
\end{equation*}
used in Heisenberg time evolution (eq. \ref{eq: Koopman_Heisenberg_equation_of_motion}) with the anticommutation relationships (eqs. \ref{eq: dirac_anticommutators}) and coordinate Fourier transform (eq. \ref{eq: coordinate_fourier_transform}) gives us the Dirac equation after evaluation:

\begin{equation}
   \Big[ i\hbar\gamma^\mu\boldsymbol{\partial}_\mu -e\gamma^\mu A_\mu(\boldsymbol{x}) - m\Big]\Psi(\boldsymbol{x})=0.
\end{equation}
Following the classical wavefunction treatment of \citet{schonberg_1,schonberg_3,schonberg_2}, we can propose anticommutators of Koopman spinors of the form 

\begin{equation}\label{eq: Koopman_spinor_anticommutators}
    \{\Psi_\alpha(x,p),\Psi_\beta^\dagger(x',p')\}=\delta_{\alpha\beta}\delta^{(3)}(x-x')\delta^{(3)}(p-p'),
\end{equation}
\begin{equation}
    \{\Psi_\alpha^\dagger(x,p),\Psi_\beta^\dagger(x',p')\}=\{\Psi_\alpha(x,p),\Psi_\beta(x',p')\}=0.
\end{equation}
Note that the phase space Fourier transform lets us re-express the anticommutators in terms of Dirac delta functions of $\lambda_x$, $\lambda_p$, etc.
We note that these anticommutators do not contradict classical mechanics, since this essentially places raising and lowering operations in the (unobservable) ``hidden dimensions" of Koopman theory \citep{schonberg_1,schonberg_3,schonberg_2}. In other words, classical observables retain their continuous spectra, and the second quantization of Koopman fields \textit{on phase space} does not quantize classical values. By example in the nonrelativistic limit, one can think of the Heisenberg uncertainty between conjugate pairs such as $x,\lambda_x$ or $p,\lambda_p$ \citep{PiaseckiThesis}, but since $\lambda_x$ and $\lambda_p$ are hidden dynamical variables, they do not affect the measurements of classical observables of the form $O(x,p)$.

With a Koopman generator of the following form,

\begin{equation}\label{eq: spinor_LK_generator}
    K = \int d^3x~d^3p ~\Psi^\dagger(x,p)\Big[i\gamma^0\gamma^j\partial_j-ie\gamma^0 \gamma^\mu  F_{\mu \nu}\frac{\partial}{\partial p^\nu}\Big]\Psi(x,p),
\end{equation}
with Heisenberg time evolution (eq. \ref{eq: Koopman_Heisenberg_equation_of_motion}), the anticommutation relationships (eqs. \ref{eq: Koopman_spinor_anticommutators}) and phase space preserving Fourier transform (eqs. \ref{eq: phase_space_Fourier}, \ref{eq: phase_space_Fourier_2}), we get the classical time evolution (compare to eqs. \ref{eq: kvn_rel_sandwich}, \ref{eq: EVM_eq}, \ref{eq: wigner_flow})
\begin{equation}
 \Big[i\gamma^\mu \partial_\mu^x+ie\gamma^\mu F_{\mu \nu}\partial^\nu_p\Big]\Psi(x,p)=0.
\end{equation}
If we multiply the above expression from the left by spacetime momentum vector $p=p_\nu \gamma^\nu$ where here $\gamma^\nu$ can be thought of as a spacetime basis vector \citep{CAS_doran_book,CAS_Spacetime_EM,Hestenes_book_spinors}, taking the rank-0 (grade-0) elements, we recover the well-known relativistic Liouville flow \citep{rel_kinetic_theory_2013}:

\begin{equation}\label{eq: Vlasov_Koopman_theory}
    \Big[ip^\mu \partial_\mu^x+iep^\mu F_{\mu \nu}\partial^\nu_p\Big]\Psi(x,p)=0.
\end{equation}
If we interpret $\rho(x,p)=\Psi^\dagger\Psi$ as particle number density, we can derive the Vlasov equation, as before.
We will discuss further the meaning and interpretation of the above Koopman spinor equations in our subsequent work. 


To construct a quantum-to-classical interpolation, we require a spinor defined on relativistic phase space that reproduces quantum behavior in one limit, including nonlocal wavenature corrections, and reduces to the Koopman spinor in the other (analogous to the spinor in eq. \ref{eq: u_v_exp}). The standard Dirac spinor, however, is defined only on coordinate space (or its Fourier dual) and therefore does not naturally extend to phase space. To build such an interpolating object, we begin with the relativistic von Neumann equation introduced in the previous section (eqs. \ref{eq: rel_density_operator}-\ref{eq: dirac-density}). As discussed previously, imposing an idempotent condition removes the redundancies of the Dirac density matrix and yields a relativistic spinor defined on phase space. Applying this construction to the von Neumann equation (eq. \ref{eq: u_v_exp}) then produces a phase space spinorial equation that captures both quantum and classical dynamics in the appropriate limits. We interpret $\Psi(x,p)$ as a quantum spinor projected to a point of classical phase space.

Now, we can work with a quantum-classical phase space spinor $\Psi(x,p)$ with the interpolating generator:

\begin{equation}
    \Omega(\hbar) = \int d^3x~d^3\lambda_p ~\Psi^\dagger(x,\lambda_p)\Big[i\gamma^0\gamma^j\partial_j-\frac{e\gamma^0\gamma^\epsilon}{\hbar}\Big(A_\epsilon(x-\hbar\lambda_p/2)-A_\epsilon(x+\hbar\lambda_p/2)\Big)\Big]\Psi(x,\lambda_p).
\end{equation}
In the classical limit, utilizing the phase space Fourier transform and gauge invariance, we recover the Koopman generator (eq. \ref{eq: spinor_LK_generator}). With Heisenberg time evolution and the phase space second quantization once again, we recover the von Neumann dynamics for a phase space spinor (compare to eq. \ref{eq: u_v_exp}):

\begin{equation}
\Big[i\hbar\gamma^\mu\partial_\mu+e\gamma^\epsilon\Big(A_\epsilon(x-\hbar\lambda_p/2)-A_\epsilon(x+\hbar\lambda_p/2)\Big)\Big]\Psi(x,\lambda_p)=0.
\end{equation}
The Wigner operator can be derived again as a phase space amplitude by means of the phase space Fourier transform of the Dirac density matrix, keeping in mind $\Psi(x,p)\cong \hat W(x,p)$. In this context, adopting the phase space second quantization framework suggested by \citet{schonberg_1, schonberg_3, schonberg_2} provides a consistent second quantized Koopman approach.

\section{Relativistic Koopman Treatment of Classical Chromodynamics}\label{nonabelian_section}

The classical limit for QCD is the high temperature quark-gluon plasma, where quarks and gluons can flow but lose much of their wavenature like in the confined quantum case \citep{Wong_eq_1970,Wong_eq_2002_2,covariant_wigner_QCD_2021}. One can conceptualize the classical limit of QCD as being typified by color degrees of freedom associated  with point particles on a well-defined phase space, interacting only through a classical Yang-Mills field \citep{chromohydro_heinz_1985,Wong_eq_2002_2}. 


Such a plasma would obey what is termed the Wong-Vlasov equation:

\begin{equation}
        \Big(P^\nu \frac{\partial}{\partial X_\nu} + eP_\iota Q^aF^{\sigma\iota}_a\frac{\partial}{\partial P_\sigma}-eP^\mu f_{abc}A^b_\mu Q^c\frac{\partial}{\partial Q^a}\Big)\rho(X^\mu,P^\nu,Q) = 0
\end{equation}
where $Q^a$ ($a=1,..,8$) are the charges associated with the generators $T^a$ of the QCD $SU(3)$ color group. In the Wong-Vlasov approach, we extend the phase space to not only include the relativistic coordinates $x^\mu,p_\nu$, but also the values of charges $Q^a$, to uniquely specify the system. Extending the phase space is one possible way to carry out the Koopman analysis, in which case we would be dealing with scalar amplitudes of the form $\psi(x,p,Q)$ \citep{Wong_eq_1970,Quark_Distribution_9,chromohydro_heinz_1985}. However, we will instead opt for a different approach similar to the analysis of the previous sections, where we take color matrix valued Wigner operators. 

%

%

The aesthetic advantage of the matrix approach is that the Wigner operators only have evolution equations based on the fundamental dynamical variables $x$ and $p$, with the matrix encoding the internal degrees of freedom. There is no need for partial derivatives with respect to spin, charge $Q^a$, or other extended phase space variables. It also means we retain canonical Poisson bracket structure and do not need to modify it to accommodate these additional variables \citep{chromohydro_heinz_1985}.  The correspondence between Wong-Vlasov-type scalar amplitudes and matrix Vlasov-type expressions is given by the moments \citep{Quark_Distribution_9,chromohydro_heinz_1985}

\begin{equation}
    w_0(x,p) = \int dQ ~\psi(x,p,Q),~~w_a(x,p)=\int dQ ~Q_a \psi(x,p,Q),
\end{equation}
where $w_0, w_a$ completely specify the color Wigner matrix \citep{Quark_Distribution_9}, allowing for the expansion

\begin{equation}\label{eq: wigner_color_expansion}
    \hat{W}(x,p)=w_0\hat{\mathbb{I}}+w_aT^a
\end{equation}
in color space. 

Higher color moments can be rewritten in terms of elements $w_0,w_a$ due to the QCD color algebra \citep{Quark_Distribution_9}. With this, the Wong-Vlasov equation can instead be written in terms of a color Wigner matrix on standard phase space in the $\hbar \rightarrow 0$, collisionless limit:

\begin{equation}\label{eq: color_matrix_1}
p^\mu\partial_\mu w_0-p^\mu F_{\mu\nu}^a\partial_p^\nu w_a=0,
\end{equation}

\begin{equation}\label{eq: color_matrix_2}
    p^\mu\partial_\mu w_a-p^\mu f_{amc}A^m_\mu w^c-p^\mu F_{\mu\nu}^b\partial_p^\nu w_{ab}=0
\end{equation}
where 

\begin{equation}
    w_{ab}=\frac{1}{6}\delta_{ab}w_0-\frac{1}{2}d_{abc}w_c
\end{equation}
due to color algebra \citep{chromohydro_heinz_1985}. \citet{Quark_Distribution_9,chromohydro_heinz_1985} has demonstrated the surprising fact that to derive the correct classical limit, one must preserve the non-commuting nature of the color algebra. In classical color kinetic theory, the infinite color moment hierarchy arises because of the non-Abelian commutation relations.

In this section, we continue with the second quantized approach introduced in the last section, but include color degrees of freedom for the Koopman spinor. In this classical treatment that respects the color algebra, one can work with $w_0, w_a$, and the effects of higher color moments are automatically captured through the structure constants. This means classical behavior is fully captured by eqs. \ref{eq: color_matrix_1}, \ref{eq: color_matrix_2} in place of the Wong-Vlasov equation \citep{chromohydro_heinz_1985}. 

Introducing color degrees of freedom to the Koopman spinor construction has several important consequences. The color degrees of freedom are a consequence of internal symmetries. The object $\Psi(x,p)$ now carries both a spinor index 
$\alpha$ and color index, which are treated independently. As a result, 
$\Psi(x,p)$ may be regarded as a matrix-valued spinor in color space.
This means that the standard symmetry analysis applies in the classical case. Assuming $\Psi(x,p)$ transforms in the adjoint representation, the local $SU(3)$ gauge transformations take the familiar form,

\begin{equation}
    F'_{\mu\nu}=\omega F_{\mu\nu} \omega^{-1},~~A_\mu'=\omega A_\mu \omega^{-1}+\frac{i}{g}\omega(\partial_\mu \omega^{-1}),~~D_\mu\Psi=\partial_\mu\Psi+[A_\mu,\Psi],
\end{equation}
where $\omega(x)=\exp[i\theta^a(x)T_a]$. Because we retain that the dependence of $\theta^a(x)$ is solely on spacetime coordinates, whether or not we work within a classical or quantum framework, the above expressions automatically follow from internal $SU(3)$ symmetry. The $SU(3)$ color symmetry is equivalently implemented on the classical level (with no wavenature between position and momentum).

In eq. \ref{eq: wigner_color_expansion}, we note that the color components $w_0, w_a$ are encoding physical information encoded in the even multivector $\hat W_+$, as discussed in the previous section. Since $\Psi(x,p) = \hat{W}_+\epsilon$, one can write
\begin{equation}\label{eq: adjoint_expansion}
    \Psi_\alpha = (w_0)_\alpha \hat{\mathbb{I}}\epsilon + (w_a)_\alpha T^a\epsilon
\end{equation}
where $[T_a,\epsilon]=0$ since one acts on the spinor space and one exists in the color space. The spinorial degrees of freedom are the same as the Abelian case \citep{QCD_quark_gluon_1986_HE}, letting us map from the even Clifford algebra into a spinor through idempotent \citep{Lounesto_2001,Clifford_algebra_book_w_spinors} in the non-Abelian case. 

In the classical limit, one derives the correct Vlasov flow for quarks and antiquarks. Starting from the Heisenberg time evolution, the Koopman generator, and phase space second quantization (now with color indeces), one can derive eq. \ref{eq: Vlasov_Koopman_theory} as before. We implement the $SU(3)$ gauge invariance leading to the usual minimal substitution,

\begin{equation}
    \partial_\mu^x \rightarrow D_\mu = \partial^x_\mu + [A_\mu,\cdot~],
\end{equation}
and recognize $F_{\mu\nu}$ as the Yang-Mills field. So altogether we get:

\begin{equation}\label{e: w_min_coupling}
        \Big[ip^\mu \partial_\mu+[A_\mu,\cdot~]+iep^\mu F_{\mu \nu}\partial^\nu_p\Big]\Psi(x,p)=0.
\end{equation}
Utilizing the adjoint color expansion of $\Psi$ (eq. \ref{eq: adjoint_expansion}) but dropping the spinor index for notational simplicity, we get:

\begin{equation}\label{eq: classical_gluo_1}
        ip^\mu \partial_\mu w_0\epsilon+ip^\mu \partial_\mu w_a T^a\epsilon + A_{\mu a}[T^a,T^b]w_b\epsilon+iep^\mu F_{\mu \nu a}T^a\partial^\nu_p w_0\epsilon +iep^\mu F_{\mu \nu a}T^aT^b\partial^\nu_p w_b\epsilon=0.
\end{equation}
We next utilize the identities \citep{wilson_lines}:

\begin{equation}
    [T^a,T^b]= if^{abc}T_c,
\end{equation}
\begin{equation}
    \{T^a,T^b\}=\frac{1}{3}\delta^{ab}+d^{abc}T_c,
\end{equation}

\begin{equation}
    T_aT_b=\frac{1}{2}\{T_a,T_b\}+\frac{1}{2}[T_a,T_b]=\frac{1}{2}if^{abc}T_c+\frac{1}{6}\delta^{ab}+\frac{1}{2}d^{abc}T_c.
\end{equation}
Applied to eq. \ref{eq: classical_gluo_1} and separating by scalar and generator $T^a$ parts, we recover the classical limit of QCD (eqs. \ref{eq: color_matrix_1}, \ref{eq: color_matrix_2}). 
This highlights the equivalence of a phase space second quantized Koopman theory and classical chromodynamics. 
The Wigner operator as a quantum amplitude can then be derived in the same manner as described in the previous section, but with QCD consistent gauge link \citep{wilson_lines}.

\section{Koopman Implications for Chromodynamics}

The Koopman perspective provides a unified framework for understanding the emergence of phase space distributions in QCD. In this work, we present a natural derivation of the Wigner operator by placing quantum and classical mechanics within a common mathematical language. Within both nonrelativistic and relativistic settings, the Wigner function appears as an interpolating object between classical and quantum descriptions. When color degrees of freedom are included, the classical behavior of the quark-gluon plasma is recovered for Koopman dynamics.

A central observation of this approach is that the Wigner function is more naturally interpreted not as a quasiprobability distribution, but as a quantum amplitude projected unto classical phase space. Its lack of positive definiteness is therefore not pathological, but a direct consequence of its amplitude-like character.

In QCD, the Wigner operator plays a central role in the construction of PDFs, GPDs and TMDs. We argue that, analogous to the nonrelativistic case, the Wigner operator corresponds to a mathematically well-defined quantum–classical spinor on phase space. This structure emerges upon removing redundant information from the Wigner operator through the use of an idempotent. In the classical limit, this construction reduces to a Koopman spinor satisfying relativistic Vlasov dynamics.

This analysis opens several avenues for further investigation. First, it clarifies the origin and interpretation of PDFs, GPDs, and TMDs in QCD. A proper understanding of the Wigner operator is essential for interpreting experimental observables derived from these quantities. Its interpretation as a multi-valued spinor implies that negative regions in parton distributions should not be regarded as defects, but as natural features arising from interference at the amplitude level. These regions encode nontrivial physical information and should therefore be viewed as meaningful rather than problematic. From this perspective, PDFs, GPDs, and TMDs are best understood as reduced projections of a more fundamental phase space amplitude, suggesting a refined framework for analyzing hadronic substructure.
It would be interesting to re-evaluate existing hadronic models from the perspective of this work.

Second, this framework provides a natural setting for revisiting the notion of spin in a classical context. Classical spin can be thought of as the high occupancy limit of quantum spin values, forming a continuum of values \citep{Ohanian_1986,barandes_qft_1,barandes_qft_2}. Spinors, as geometric objects encoding orientation and momentum, need not be intrinsically quantum. Instead, they may be understood as geometric structures that arise in both classical and quantum theories (e.g., \citealt{spinorial_5,Hestenes_book_spinors}). In this sense, the Koopman formalism presented here offers a pathway toward a classical description of continuous spin degrees of freedom. 

Indeed, Koopman-like spinor formulations for semiclassical electrons with spin may already be implicit in condensed matter, solid state, and plasma physics contexts \citep{KS_condensed_matter_JH_2017,KS_application_spin_current_JH_2018,KS_solid_state_plasma_GM_2019}, suggesting that the ideas developed here may have broader applicability beyond high-energy QCD. We plan to further explore the properties of Koopman spinors in an upcoming work.

\section*{Appendix I: Phase Space Fourier Transform}
To create a Hilbert phase space, we have to double the Hilbert coordinate space \citep{Bohm_KvN,wave_operator}. Eq. \ref{eq: density_to_amplitude} depicts this explicitly, where we have a coordinate space density on the right and a wavefunction on phase space on the left. The Koopman algebra (eqs. \ref{eq: classical_comm_1a}, \ref{eq: koopman_algebra}) further reflects this. We can write the Koopman algebra from a doubled symmetrized quantum algebra \citep{Bohm_KvN,wave_operator}:

\begin{equation}
    \hat{x}=\frac{1}{2}(1\otimes \boldsymbol{\hat{x}}^\dagger +\boldsymbol{\hat{x}}\otimes1),~~\hat{p}=\frac{1}{2}(\boldsymbol{\hat{p}}\otimes 1 + 1\otimes\boldsymbol{\hat{p}}^\dagger),
\end{equation}
\begin{equation}
    \hbar\hat\lambda_x = \boldsymbol{\hat{p}}\otimes 1 - 1\otimes\boldsymbol{\hat{p}}^\dagger,~~\hbar\hat\lambda_p=1\otimes \boldsymbol{\hat{x}}^\dagger -\boldsymbol{\hat{x}}\otimes1.
\end{equation}
The wavenature of nonrelativistic quantum mechanics is encoded in the canonical commutator. A simple but elegant derivation in \citet{bondar_book_2025} demonstrates the connection. Assuming a commutator of the form $[\hat{A},\hat{B}]=i\kappa$, then the Fourier transform is a consequence: 

\begin{equation}
    \bra{A'}[\hat{A},\hat{B}]\ket{A}=(A'-A)\bra{A'}\hat{B}\ket{A}=i\kappa\delta(A'-A),
\end{equation}
which implies
\begin{equation}
    \bra{A'}\hat{B}\ket{A}=i\kappa\frac{\partial}{\partial A}\delta(A'-A).
\end{equation}
So then

\begin{equation}
    \bra{A}\hat{B}\ket{B}=B\braket{A|B}=\int dA' \bra{A}\hat{B}\ket{A'}\braket{A'|B}=-i\kappa\frac{\partial}{\partial A}\braket{A|B}.
\end{equation}
Solving for $\braket{A|B}$ gives us \citep{bondar_book_2025}

\begin{equation}
    \braket{A|B}=\frac{1}{\sqrt{2\pi}}e^{iAB/\kappa}.
\end{equation}
With the canonical commutator, this implies the quantum wavenature $\braket{\boldsymbol{x}|\boldsymbol{p}}=\frac{1}{\sqrt{2\pi}}e^{i\boldsymbol{x}\boldsymbol{p}/\hbar}$. For relativistic quantum mechanics, this expression is generalized as the relationship between the four-position and the conjugate four-momentum, avoiding the issue of a self-adjoint time operator. However, with the Koopman algebra, this implies the classical ``wavenature" $\braket{x,p|\lambda_x\lambda_p}=\frac{1}{2\pi}e^{i\lambda_xx+i\lambda_pp}$. In the same way, we generalize the Fourier expression to be in terms of \textit{classical} four-position and four-momentum, giving us the phase space preserving Fourier transform, eqs. \ref{eq: phase_space_Fourier}, \ref{eq: phase_space_Fourier_2}. Alternatively, we can think of the Koopman Fourier transform as the symmetrized doubling of the quantum coordinate Fourier transform, as mentioned above for the Koopman algebra.

Thus, we can speak of classical wave amplitudes in phase space \citep{schonberg_1,schonberg_2,Riccia_KvN,Bohm_KvN,Propogator_KvN}. The classical ``wavenumbers" $\lambda_x, \lambda_p$ represent the inverse Fourier space of classical $x,p$.

\section*{Appendix II: Correspondence between phase space spinors and Wigner operator}
Here we carry out an example calculation to demonstrate explicitly how the Wigner operator maps directly into a phase space spinor. Assuming Dirac matrices in Dirac-Pauli representation:

\begin{equation}
    \gamma^0 \dot{=}\begin{bmatrix}
        1 &0&0&0\\
        0&1&0&0\\
        0&0&-1&0\\
        0&0&0&-1
    \end{bmatrix},~~\gamma^j\dot{=}\begin{bmatrix}
        0&\sigma^j\\
        -\sigma^j&0
    \end{bmatrix}.
\end{equation}
Eq. \ref{eq: second_idempotent} is a standard idempotent for Dirac spinors \citep{Lounesto_2001, CAS_idempotent_2015}. With the above matrices, the idempotent representation becomes

\begin{equation}
    \epsilon \dot{=} \begin{bmatrix}
        1 &0&0&0\\
        0 &0&0&0\\
        0 &0&0&0\\
        0 &0&0&0\\
    \end{bmatrix}.
\end{equation}
The Wigner operator under constraints has 8 real spinorial degrees of freedom, including in the QCD case \citep{QCD_quark_gluon_1986_HE}. Both Dirac spinors and the constrained Wigner operator can be expressed in terms of a scalar, a pseudoscalar, and 6 tensorial degrees of freedom \citep{hestenes_dirac_1975, QCD_quark_gluon_1986_HE}. The 8 real functions on phase space in the constrained Wigner operator maps to an even multivector,

\begin{equation}
    \hat{W}_+=\mathcal{F}+i\gamma^5 \mathcal{P}+\frac{1}{2}S_{\mu\nu}\sigma^{\mu\nu}.
\end{equation}
Note that $\sigma^{\mu\nu} = \frac{i}{2}[\gamma^\mu,\gamma^\nu]$. Using the Dirac-Pauli representation, we can write 

\begin{equation*}
    \sigma^{01}\dot{=}\begin{bmatrix}
        0&0&0&i\\
        0&0&i&0\\
        0&i&0&0\\
        i&0&0&0
    \end{bmatrix},~~\sigma^{02}\dot{=}\begin{bmatrix}
        0&0&0&1\\
        0&0&-1&0\\
        0&1&0&0\\
        -1&0&0&0
    \end{bmatrix}, ~~\sigma^{03}\dot{=}\begin{bmatrix}
        0&0&i&0\\
        0&0&0&-i\\
        i&0&0&0\\
        0&-i&0&0
    \end{bmatrix}
\end{equation*}
\begin{equation}
    \sigma^{13}\dot{=}
    \begin{bmatrix}
        0&i&0&0\\
        -i&0&0&0\\
        0&0&0&i\\
        0&0&-i&0
    \end{bmatrix},~~\sigma^{12}\dot{=}
    \begin{bmatrix}
        1&0&0&0\\
        0&-1&0&0\\
        0&0&1&0\\
        0&0&0&-1
    \end{bmatrix},
    ~~\sigma^{23}\dot{=}\begin{bmatrix}
        0&1&0&0\\
        1&0&0&0\\
        0&0&0&1\\
        0&0&1&0
    \end{bmatrix}.
\end{equation}
Then,

\begin{equation}
    \Psi(x,p) = \hat{W}_+\epsilon =\begin{bmatrix}
        \mathcal{F}+S_{12}&0&0&0\\
        S_{23}+iS_{31} &0&0&0\\
        i\mathcal{P}+iS_{03} &0&0&0\\
        S_{20}+iS_{01}&0&0&0 
    \end{bmatrix}~~\leftrightharpoons~~\Psi(x,p)=\begin{bmatrix}
        \mathcal{F}+iS_{12}\\
        S_{23}+iS_{31}\\
        \mathcal{P}+iS_{03}\\
        S_{20}+iS_{01} 
    \end{bmatrix}.
\end{equation}
Dirac spinor has the same algebraic structure as the quantity on the left \citep{hestenes_dirac_1975}, just is a function of coordinate space or its conjugate momentum instead of existing on phase space. This mapping can also be done clearly within the Clifford Spacetime Algebra formalism \citep{spinorial_2,Hestenes_book_spinors}. The 8 real degrees of freedom of the constrained Wigner operator get mapped into a spinor on phase space. 
{\newline

\noindent\rule{\linewidth}{0.5mm} 
\vspace{2mm}
\begin{tikzcd}[row sep=3em, column sep=5em]
& 
\hat{W}_+ \in Cl^+_{(1,3)}
\arrow[dl, "\left(1+\frac{\gamma^\mu p_\mu}{m}\right)\cdot"']
\arrow[dr, "\cdot\,\epsilon"]
& 
\\
\hat{W} \in Cl_{(1,3)}
& 
& 
\Psi(x,p) \in \mathbb{C}^4
\end{tikzcd}
}
\newline
\textbf{Figure 1:} Relationship between spinor operator $\hat W_+$, Wigner operator $\hat W$ in the limit of $\hbar \rightarrow 0$ (quantum limit would have a similar projector, but include quantum corrections in terms of $\hbar$), and phase space Koopman spinor $\Psi(x,p)$. The same 8 real degrees of freedom exist in different Clifford algebra representations. 

\section*{Acknowledgements} We would like to thank Denys Bondar of Tulane University for many indepth discussions and feedback on the Koopman formulation. This work was supported in part by the U.S. Department of Energy (Grant No. DE-FG02-
03ER41260). The National Energy Research Scientific Computing Center (NERSC) supported by the Office of Science
of the U.S. Department of Energy under Contract No. DE-AC02-05CH11231 is also acknowledged.


%

%

%

%




\bibliographystyle{apalike}
\bibliography{hilbert_space}
\end{document}